\documentstyle[12pt]{article}
\begin{document}

\begin{center}
{\bf Neutron matter with a model interaction} \bigskip \bigskip

M. Ya. Amusia$^{a,b}$, V. R. Shaginyan$^{c,d}$
\footnote{E--mail: vrshag@thd.pnpi.spb.ru}\bigskip

$^{a}$The Racah Institute of Physics, Hebrew University, Jerusalem 91904,
Israel;\\[0pt]
$^{b}$A.F. Ioffe Physical-Technical Institute, 194021 St.Petersburg,
Russia; \\[0pt]
$^{c}$Petersburg Institute of Nuclear Physics, 188350 Gatchina,
Russia;\\[0pt] $^{d}$ Department of Physics, University of
Washington, Seattle, WA 98195, USA \end{center}

\begin{abstract}\noindent
An infinite system of neutrons interacting by a model pair potential
is considered. We investigate a case when this potential is
sufficiently strong attractive, so that its scattering length $a$
tends to infinity, $ a\rightarrow -\infty $. It appeared, that if
the structure of the potential is simple enough, including no finite
parameters, reliable evidences can be presented that such a system is
completely unstable at any finite density. The incompressibility as a
function of the density is negative, reaching zero value when the
density tends to zero. If the potential contains a sufficiently
strong repulsive core then the system possesses an equilibrium
density. The main features of a theory describing such systems are
considered.
\end{abstract}

\vspace{0.8cm} \noindent {\bf PACS}. 24.10.Cn Many-body theory -
21.10.Dr Binding energies

\vspace{0.8cm}
\noindent
There exists a well known problem in many body physics related to the
description of the ground state properties of an infinite system
composed of interacting fermions. In general, this description is
based usually on tedious numerical calculations, particularly when
the interaction is rather strong. The well known exceptions from this
situation, when it is possible to calculate the ground state
properties analytically, are the Random Phase Approximation (RPA) for
a high density electron gas \cite{gb} and the low density
approximation for dilute gases \cite{ly}. In both cases the kinetic
energy $T_{k}$ is much bigger then the interaction energy $E_{int}$
of the system. This allows to apply some kind of a perturbation
theory. In the case of homogeneous electron liquid it turns out that
the analytical RPA-like description is also possible not only at very
high but medium densities when $T_{k}\sim E_{int}$ \cite{sh,as}.
Similar extension of the range of validity is impossible in the case
of fermion systems at low densities $\rho $: there the gas
approximation is not applicable if $T_{k}\sim E_{int}$. In the cases
when the pair interaction is attractive and sufficiently strong, the
system can have a quasi equilibrium or equilibrium states in which
$T_{k}\simeq E_{int}$. On the other hand, these states are preceded
by special points with density $\rho $ values at which the
incompressibility $K(\rho )$ of the system tends to zero. Thus, if it
would be possible to predict the existence of such points then in
principle it would become possible to conclude that the system has at
least a quasi equilibrium state.

In this Short note we address the above mentioned problem and consider
the ground state properties of the infinitely extended multi-fermion
system. We demonstrate that it can be done analytically provided that
the pair interaction between fermions is characterized only by the
scattering length $ a\rightarrow -\infty$. One can say in this case
that the scattering length is the dominant parameter of the problem
under consideration. Such an investigation is of great importance
since it can be applied to fermion systems interacting via potentials
with not only infinite, but also sufficiently large $a$. For
instance, the scattering length $a$ of neutron-neutron interaction is
about $-20$ fm, thus greatly surpassing the radius of the interaction
$r_{0}$. On the other hand, it is possible now to prepare
artificially a system composed of Fermi atoms interacting by an
artificially constructed potential with almost any desirable scattering
length, similarly to that how it is done for the trapped Bose gases, see
e.g. \cite{nat}. An experimental study, performed on such
Fermi-system would be of great importance presenting new information
on the behavior of dilute gases and the gas-liquid phase transition.

Let us consider the interaction of two isolated particles. We assume
that this interaction is of finite radius $r_{0}$, which is small, so
that $ p_{F}r_{0}\ll 1$ ($p_{F}=(3\pi ^{2}\rho )^{1/3}$ is the Fermi
momentum), but its strength is such that the scattering length is
negative and infinitely big, $a\rightarrow -\infty$. We assume also,
that the density $\rho $ of the system under consideration is
homogeneous. As it will be demonstrated below, in such a case the
system is located in the vicinity of a phase transition, transforming
it into a strongly correlated one. Therefore, the problem of
calculating its ground state properties has to be treated for the
most part qualitatively.

Let us start considering general properties of a Fermi system with
some attractive two-particle bare interaction $V(r)$, which is
sufficiently weak to create a two-particle bound state. Assume, that
the scattering length $a$ corresponding to this potential is negative
and finite. The ground state energy density $E(\rho )$ can be
approximated by Skyrme-like expression \cite {vb},
\begin{equation}
E(\rho )=\frac{3p_{F}^{2}}{10M}\rho +t_{0}\rho ^{2}+t_{3}\rho ^{7/3},
\end{equation}
Here $M$ is the particle mass, and $\rho $ is their density in the
system.  The first term of eq. (1) is the kinetic energy $T_{k}$,
while the second and the third are related to the interaction energy
$E_{int}$ determined by the potential $V(r)$. The second term which
is proportional to $t_{0}$ gives a proper description in the gas
limit. The third term provides the behavior of $E(\rho )$ at higher
densities, including that of the equilibrium density. Such structure
of $E(\rho )$ appears if the interaction is sufficiently attractive
so that $t_{0}<0$, and $t_{3}>0$. Note, that eq. (1) presents at
least a qualitative description of the system under consideration
giving a rather simple and reasonable presentation of the function
$E(\rho )$. A more precise picture of the energy dependence upon
density can be obtained using more sophisticated functionals for the
ground state energy \cite{fa}.

If the potential $V$ is of short range and purely attractive, then
in the Hartree-Fock approximation the ground state energy $E_{HF}$ is
given by the following expression \begin{equation} E_{HF}(\rho
)=\frac{3p_{F}^{2}}{10M}\rho +t_{HF}\rho ^{2}, \end{equation} where
the parameter $t_{0}=t_{HF}$ , being negative, is entirely determined
by the potential $V(r)$. For instance, in the case of a short range
$\delta$-type interaction one has $t_{HF}=-v_{0}/4$, with $v_{0}$
being the corresponding strength of the potential. Eq. (2) shows that
at small densities $E_{HF}>0$ due to the kinetic energy term, but at
sufficiently high densities $\rho \rightarrow \infty $ the
Hartree-Fock energy becomes dominating, leading to the collapse of
the system, with $E_{HF}\rightarrow -\infty $. Keeping in mind that
the Hartree-Fock approximation gives the upper limit to the binding
energy $E_{HF}\geq E$, one can conclude that the system does not
have, in this case, an equilibrium density $\rho_{e}$ and energy
$E_{e}$ since $E_{e}\rightarrow -\infty $ when $\rho\rightarrow
\infty $ \cite{ll}. Note, that for a given and finite total number of
particles $N$, the HF energy is not going to infinity and the system
collapses into a small volume with the radius $r_{0}$, with the
density $\rho\sim N/r_{0}^{3}$. It is evident that the function
$E(\rho)$ is positive at small densities, if the parameter $t_{0}$ is
finite. Therefore, it must have at least one maximum at the density
$\rho_{m}$ before it becomes negative, on the way to
$E\rightarrow-\infty $. If the potential $ V(r)$ includes a
repulsive core at sufficiently short distances, then $t_3>0$ . As a
result, the system has an equilibrium density and energy, $\rho _{e}$
and $E_{e}$, respectively, determined by the repulsive core strength
and its radius $r_{c}\sim r_{0}$.

Now let us apply eq. (1) to demonstrate the most important features
of the system under consideration:

a) when $\rho \rightarrow 0$ the third term on r.h.s. in eq. (1) can
be omitted. The kinetic energy is relatively very big,
$T_{k}\gg E_{int}$, and $ t_{0}\sim a$, with $a<0$ being the
scattering length. In that case we have a dilute Fermi gas with
positive pressure $P$ and incompressibility $K$, the latter being
determined by the equation, see e.g. \cite{pn}, \begin{equation}
K(\rho )=\rho^{2} \frac{dE^{2}(\rho)}{d\rho^{2}}.
\end{equation}

b) on the way to higher densities, which can be achieved by applying
an external pressure, the system
reaches the density $\rho _{c1}<\rho_{m}$ at which the
incompressibility is equal to zero, $K(\rho _{c1})=0$. Remembering that
at the maximum the second derivative is negative, one can conclude,
as it is seen from eq. (3), that $K(\rho _{m})<0$. In the range $\rho
_{c2}\geq \rho \geq \rho _{c1}$ the incompressibility is negative,
$K<0$, and as a result the system becomes totally unstable. In fact,
in this density range all calculations of the ground state energy are
meaningless since such a system cannot exist and thus be observed
experimentally \cite{lanl};

c) at some point $\rho=\rho_{c2}>\rho_m$ the contribution due to the
repulsive core becomes sufficiently strong to prevent the further
collapse of the system.
The incompressibility attains $K=0$ at $\rho_{c2}<\rho_{e}$, being
positive at the higher densities. Finally, the system becomes stable
at $\rho>\rho_{c2}$, reaching equilibrium density at $\rho_{e}$ with
equilibrium energy equal to $E_{e}$. It is obvious that
$K(\rho_{e})>0$ being proportional to the second derivative at the
minimum, see eq. (3). It should be kept in mind that in this density
domain, $\rho \geq \rho_{c2}$, the function $E(\rho)$ is determined
by the repulsive part of the potential which makes $t_3>0$. As it was
mentioned above, without this component of $ V(r)$ the system's
energy would infinitely increase, $E_{HF}\rightarrow -\infty $, with
density growth, $\rho \rightarrow\infty$, thus inevitably collapsing.

One could expect in principle the existence of metastable states at
$\rho>\rho_{c1}$ if the potential $V(r),$ even being pure
attractive, would have a complicated structure. It can be said that
there could exist parameters of $V(r)$, which are able to open the
possibility for the metastable states to be formed. On the other
hand, a system of fermions interacting via a short-range, finite
scattering-length, $\delta $-type potential $V_{\delta }$ , can be
stable only in the dilute gas regime. While at the densities $\rho
\geq \rho _{c1}$ the incompressibility $K$ becomes negative, the system
collapses. Indeed, the potential $V_{\delta }$ has no structure to
ensure any metastable states at the densities $\rho \geq \rho _{c1}$.
As a result, one can write down a dimensionless expression for the
ground state energy as a function of the only variable $z=p_{F}a$
\cite {ly,bk}, \begin{equation} \alpha E(z)=z^{5}(1+\beta (z)),
\end{equation}
with $\alpha=10\pi ^{2}Ma^{5}$. In the low density limit,
$|ap_{F}|\ll 1$ and when the interaction has the radius $r_{0}$, eq.
(4) reads \cite{bk},
\begin{equation} \alpha E(z)=z^{5}\left[
1+\frac{10}{9\pi }z+ \frac{4}{21\pi ^{2}}
(11-2\ln 2)z^{2}+\left(\frac{r_{0}}{a}\right)^{3}z^{3}
\gamma\left(\frac{r_{0}}{a}
,z\right) +...\right] .
\end{equation}
Here the function $\gamma (y,z)$ is of the order of one, $\gamma
(y,z)\sim 1$. It is seen from eq. (5) that as soon as the
scattering length becomes big enough, $|a|\gg r_{0},$ one can omit the
contribution coming from the function $\gamma $ and neglect all the
term proportional to $(r_{0}/a)^{3}$. Then eq.  (5) reduces to eq.
(4). Thus, in the case $|a|\rightarrow \infty $ we can use eq. (4) to
determine the ground state energy $E$. Eq. (4) is valid up to the
density $\rho _{c1}$ which is a singular point of
the function $\beta(z)$, since beyond this point $K<0$, and the
system is completely unstable. On the other hand, there is no
physical reasons to have another irregular point in the region $0\leq
\rho \leq \rho _{c1}$. We note, that eq. (5) is as well valid up
to its own density $\rho^{'}_{c1}\simeq\rho_{c1}$, provided
$|a|\gg r_0$.  Using eq. (3) for the incompressibility, one can
calculate the position of the point $z_{c1}$ where $K=0$.  Denoting
the corresponding $z$ as $z_{c1}=c_{0}$, where $c_{0}$ is a
dimensionless number, one is led to the conclusion that $\rho
_{c1}\sim |a|^{-3}$ provided $a$ is sufficiently large to be the only
dominating parameter. The system has only one stable region at small
densities $\rho \leq \rho _{c1}$ which decreases and even vanishes as
soon as $a\rightarrow -\infty $.  One could expect that
$|c_{0}|\rightarrow \infty $ as soon as $a$ becomes the dominant
parameter so that the above given expression for $ \beta(z)$ is valid
in the whole domain $|z|\leq\infty$. On the other hand, there exists
another singular point $z_{c2}$ in the function $E(\rho)$ and the
position of this point which corresponds to $\rho_{c2}$ depends
mainly on the parameters such as $r_{0},$ core radius $r_{c}$ of
$V(r)$ rather than on the scattering length $a$. As to the function
$\beta(z),$ by definition it does not contain any information about
$\rho_{c2}$.  Therefore, in order to take into account the existence
of $\rho_{c2}$, one has to recognize that $c_{0}$ is finite, and the
densities $\rho_{c1}$ and $\rho_{c2}$ are different. As a result, the
function $\beta $ is determined in fact only in the region $|z|\leq
|z_{c1}|$.

Now let us consider the calculation of the ground state energy $E$
of the system when the density approaches
$\rho \rightarrow \rho_{c1}\sim |a|^{-3}$
from the low density side. As a rule, points
$\rho _{c1}$ and $\rho _{c2}$ are missed in calculations because of
the lack of the self consistency \cite {jls,j,ksk}, which relates the
linear response function of system with its incompressibility $K$,
\begin{equation}
\chi (q\rightarrow 0,i\omega \rightarrow 0)
=-\left( \frac{d^{2}E}{d\rho ^{2}}
\right) ^{-1}.
\end{equation}
As we shall see below, these points can give important contributions
to the ground state energy. To see it we express the energy of a
system in the following form (see e.g. \cite{pn}), \begin{equation}
E(\rho )=T_{k}+E_{H}-\frac{1}{2}\int
\left[ \chi (q,i\omega ,g)+2\pi \rho
\delta (\omega )\right] v(q)
\frac{d{\bf q}\,d\omega \,dg}{g(2\pi )^{4}},
\end{equation}
where $\chi (q,i\omega ,g)$ is the linear response function on the
imaginary axis and $v(q)=gV(q)$, with $V(q)$ being the Fourier image
of $V(r)$. The integration over $\omega $ goes from $-\infty $ to
$+\infty $, while the integration over the coupling constant $g$ runs
from zero to the real value of the coupling constant, i.e. to $g=1$.
At the point $\rho =\rho _{c1}$ the linear response function has a
pole at the origin of coordinates $ q=0,\,\omega =0$ due to eq. (6).
At the densities $\rho >\rho _{c1}$ the function $\chi (q,i\omega )$
has poles at finite values of the momentum $q$ and frequencies
$i\omega $. This prevents the integration over $i\omega $, making the
integral in eq. (7) divergent. Thus, we conclude that it is the
contribution of these poles that reflects the system's instability
in the density range $\rho _{c1}\leq \rho \leq \rho _{c2}$. Note,
that violations of eq. (6) lead to serious errors in the calculation
of the ground state energy. Eq. (7) can be rewritten, explicitly
accounting for the effective interparticle interaction $R(q,i\omega
,g)$, (see e.g. \cite{ksk}), in the following form
\begin{equation}
E(\rho )=T_{k}+E_{H}-\frac{1}{2}\int
\left[ \frac{\chi _{0}(q,i\omega )}
{1-R(q,i\omega ,g)\chi _{0}(q,i\omega )}
+2\pi \rho \delta (\omega )\right]
v(q)\frac{d{\bf q}\,d\omega \,dg}{g(2\pi )^{4}}.
\end{equation}
Here $\chi _{0}$ is the linear response function of noninteracting
particles, while $\chi $ is given by the following equation
\begin{equation}
\chi (q,\omega )=\frac{\chi _{0}(q,\omega )}{1-R(q,\omega )\chi
_{0}(q,\omega )}.
\end{equation}
It is seen from eqs. (6) and (9)
that the denominator $(1-R\chi_{0})$ vanishes at
$\rho \rightarrow \rho _{c1}$ while the radius of
correlation tends to infinity \cite{lanl}. Thus, it is impossible to
present the denominator as a power series in $R\chi _{0}$
approximating the expansion by the finite number of terms. This
result is quite obvious since $\rho _{c1}$ is a singular point  in
the function $E(\rho )$ \ which makes it impossible to expand that
function in the vicinity of this point. Therefore, one should try to
satisfy eq. (6) in order to get proper results for the ground state
calculations in the vicinity of the instability points. Such an
approach was suggested in \cite{sh,as,ksk} and is based on the exact
functional equation for the effective interaction $R(q,\omega ,g)$,
\begin{equation}
R(q,\omega ,g_{0})=g_0v(q)-\frac{1}{2}\frac{\delta ^{2}}{\delta \rho
^{2}(q,\omega )}\int \frac{\chi _{0}(k,iw)}
{1-R(k,iw,g)\chi _{0}(k,iw)}\,v(q)
\frac{d{\bf k}\,dw\,dg}{g(2\pi )^{4}}.
\end{equation}
As a result, the linear response function $\chi $ given by eq. (9)
automatically satisfies eq. (6) \cite{ksk}. Our preliminary
calculations \cite{ksk,ksh}, based on eq. (9) and applied to the case
when the scattering length is sufficiently large but finite, confirm
the result that $\rho _{c1}\sim |a|^{-3}$.

Let us suppose for a while that the bare potential is pure
attractive. Then, the interval of the densities $[0,\rho _{c1}]$
within which the system is stable vanishes with the growth of $|a|$.
As a result, in the limit $ a=-\infty $ the incompressibility
becomes negative $K\leq 0$, making the considered system completely
unstable at any density. Thus, the point at which $a=-\infty $ is the
only point of the system's instability at all the densities. As soon
as the scattering length deviates from its infinite value, that is
$+\infty >a\ >-\infty $ the system comes back to its stable state at
list in the range of the density values $\rho \ <\rho _{c1}\sim
|a|^{-3}$. It is of interest to understand whether it is possible to
prove by e.g. numerical calculations, that $\rho _{c1}\sim |a|^{-3}$
when $ a\rightarrow -\infty $. From our point of view, at least at
this moment, the answer is ``no''. We are dealing with a system
located in the vicinity of a phase transition, which transforms it
into a strongly correlated one. As a result, it is hard to believe
that the numerical calculations could be reliable. On the other hand,
it is not really necessary to carry out numerical calculations if the
problem allows a qualitative analysis. It was argued above, that
there exists the only parameter to characterize the system which is
the scattering length $a$. In fact the scattering length determines
only the specific point $\rho _{c1}$ at which the incompressibility
vanishes, separating the region of a dilute gas from the region of the
system's instability. As soon as $a\rightarrow -\infty $ this last
and the only parameters vanishes, driving the point $\rho _{c1}\sim
|a|^{-3}$ of the curve $E(\rho )$ to the origin of coordinates.
Thereafter, the system becomes unstable at all densities. And vice
versa, as soon as the scattering length becomes finite the system is
stable at list within the interval $\rho \leq \rho _{c1}\sim
|a|^{-3}$.

Note, that as it follows from our consideration, any Fermi system
possesses an equilibrium density and energy if the bare
particle-particle interaction contains a repulsive core and its
attractive part is strong enough, so that $a\rightarrow-\infty$.
Indeed, at sufficiently small densities the ground state energy is
negative (since the incompressibility $K\leq 0)$ and the system will
collapse until the core stops the density growth. Therefore, the
minimal value of the ground state energy must be negative when the
repulsive core will enter the play to prevent the system from the
further collapse. It is worth to remark, that superfluid correlations
cannot stop the system squeezing, since their contribution to the
ground state energy being negative increases in the absolute value
with the growth of the density.

A liquid similar to the model one considered in this paper exists in
Nature.  This is liquid $^{3}He$. If a helium dimer exists, its bound
energy does not exceed $10^{-4}$ meV while the ground state energy of
helium liquid is about $2\ast 10^{-1}$ meV per atom \cite{elg}.
Because of this huge difference in binding energies, it is evident
that there is no essential contribution coming from the binding
energy of the dimer to the ground state energy of the liquid. In
fact, the numerical calculations show that the pair potential is
rather weak to produce the dimer $He_{2}$ \cite{elg}. Thus, one can
reliably consider an infinite homogeneous system of Helium atoms as
consisting of particles interacting via pair potential,
characterized by a very big but finite scattering length
$|a|\gg r_{0}$. Let us make also the following additional remark. It
seems quite probable that the neutron-neutron scattering length
($a\simeq -20$ fm) is sufficiently large to permit the neutron
matter to have an equilibrium energy and density \cite {ksh}.
Therefore, calculations of a neutron matter satisfying eq. (7) are
quite desirable.

In summary, the homogeneous system of interacting fermions was
considered.  It was shown that when the scattering length $a$ is
negative and sufficiently large the fermion matter becomes a strongly
correlated system at the densities $\rho \sim |a|^{-3}$. Therefore,
the consideration of such a system is connected to a number of
problems which yet persist and have to be resolved. At the same time,
the qualitative consideration presented above gives strong evidences
that the point $\rho _{c1}$ at which the incompressibility vanishes is
defined by $\rho _{c1}\sim |a|^{-3}$ provided the scattering length
is the dominant parameter of the problem. Thus, a homogeneous system
composed of fermions, interacting via a pure attractive potential, at
$a\rightarrow -\infty $ is completely unstable at all the densities,
with the incompressibility as a function of the density being always
negative. As soon as the density $\rho $ goes to zero the
incompressibility goes to zero as well.

We thank G.F. Bertsch for attracting our attention to the discussed
above many body problem which he has in fact suggested \cite{gfb}.
One of us (VRS) is grateful to A. Bulgac for valuable discussions,
and to the Department of Physics of the University of Washington
where part of this work was done, for hospitality. This research was
funded in part by INTAS under Grant No.  INTAS-OPEN-97-603.

\end{document}